\documentclass[twocolumn,showpacs,amsmath,amssymb,pra]{revtex4}
\usepackage{amsxtra}
\usepackage{amssymb}
\usepackage{amsmath}
\usepackage{graphicx}

\begin{document}
\title{Nonlocal dispersion cancellation with phase-sensitive Gaussian-state light}
\date{\today}
\author{Jeffrey H. Shapiro}
\affiliation{Research Laboratory of Electronics, Massachusetts Institute of Technology, Cambridge, Massachusetts 02139, USA}
  
\begin{abstract} 
Franson's paradigm for nonlocal dispersion cancellation [J. D. Franson, Phys. Rev. A {\bf 45,} 3126 (1992)] is studied using two kinds of jointly Gaussian-state signal and reference beams with phase-sensitive cross correlations.  The first joint signal-reference state is nonclassical, with a phase-sensitive cross correlation that is at the ultimate quantum-mechanical limit.  It models the outputs obtained from continuous-wave spontaneous parametric downconversion.  The second joint signal-reference state is classical---it has a proper $P$ representation---with a phase-sensitive cross correlation that is at the limit set by classical physics.  Using these states we show that a version of Franson's nonlocal dispersion cancellation configuration has essentially identical quantum and classical explanations \em except\/\rm\ for the contrast obtained, which is much higher in the quantum case than it is in the classical case.  This work bears on Franson's recent paper [J. D. Franson, arXiv:0907:5196 [quant-ph]], which asserts that there is no classical explanation for all the features seen in quantum nonlocal dispersion cancellation. 
\end{abstract}
\pacs{03.67.Mn, 42.50.Ar, 42.50.Dv}  

\maketitle

\section{Introduction}
Nonlinear interactions in $\chi^{(2)}$ materials have long been used to produce nonclassical light, including optical parametric amplifier sources of squeezed states \cite{squeezing}, optical parametric oscillator sources of photon twin beams \cite{twins}, and spontaneous parametric downconversion sources of polarization-entangled photon pairs \cite{entanglement}.  Whereas Gaussian-state quadrature statistics are invariably employed to understand the behavior of squeezed states and twin beams, the biphoton state is commonly used to describe downconverter experiments that employ coincidence counting and post-selection.  Yet, as shown in \cite{ShapiroSun, ShapiroQGN, Erkmen}, there is a unified Gaussian-state analysis capable of treating all of these nonclassical phenomena and more, e.g., the dispersion cancellation experiment of Steinberg \em et al\/\rm.\@ \cite{Steinberg} and the ghost imaging experiment of Pittman \em et al\/\rm.\@ \cite{Pittman}, both of which relied on biphoton explanations.  Recently, Franson \cite{Franson1} has argued that his dispersion cancellation paradigm \cite{Franson2} differs from that of Steinberg \em et al\/\rm.\@ in that the former is nonlocal whereas the latter is not.  More importantly, in \cite{Franson1} Franson reviews various classical strawmen that have been suggested as providing explanations for nonlocal dispersion cancellation, and shows that each of them fails to reproduce one or more of the major features of quantum nonlocal dispersion cancellation.  Hence, he concludes that nonlocal dispersion cancellation is a fundamentally quantum effect akin to violation of Bell's inequality.  

The list of classical strawmen that Franson considers does \em not\/\rm\ include the classical Gaussian state that most closely resembles the nonclassical Gaussian state emitted by a continuous-wave downconverter.  In this paper we will rectify that omission, and show that the key feature of nonclassical-state nonlocal dispersion cancellation that is not reproduced by this classical counterpart is the high-contrast nature of the photocurrent cross-correlation pattern.  This result is in keeping with what we have previously established \cite{ShapiroSun} for the Steinberg \em et al\/\rm.\@ experiment and for a similar comparison between classical-state and nonclassical-state ghost imaging, \cite{Erkmen}.  In essence, we will see that nonlocal dispersion cancellation is a consequence of classical-physics propagation of the phase-sensitive cross correlation between the signal and reference beams through the dispersive elements, but the observability of the effect  is greatly enhanced by the use of nonclassical light.  

The rest of the paper is organized as follows.  In Sec.~II we describe the measurement configuration to be analyzed.  In Sec.~III we derive the ensemble-average photocurrent cross correlation for the Sec.~II apparatus when its signal and reference beams are in a zero-mean, continuous-wave, jointly Gaussian state whose baseband field operators have phase-insensitive autocorrelations and a phase-sensitive cross correlation, but no phase-sensitive autocorrelations or phase-insensitive cross correlation.  Depending on the strength of the phase-sensitive cross correlation in comparison with the phase-insensitive autocorrelations, this state could be classical, i.e., a classically-random mixture of coherent states for which the joint density operator has a proper $P$ representation and semiclassical photodetection may be employed \cite{footnote}.  Alternatively, it could be a nonclassical state, for which no proper $P$ representation exists and quantum photodetection is required to properly analyze the measurement statistics \cite{Erkmen}.  Thus, in Sec.~IV, we exhibit the consequences of this dichotomy by evaluating our photocurrent cross correlation from Sec.~III when the joint state of the input beams either has a phase-sensitive cross correlation that is at the ultimate quantum limit, or that cross correlation saturates the tighter bound associated with classical physics.  Here we shall see that nonlocal dispersion cancellation occurs with both the nonclassical and classical states, but their contrasts differ dramatically.  In Sec.~V we close with some concluding discussion, which includes connecting our Gaussian-state analysis to the more frequently employed biphoton treatment of dispersion cancellation.

\section{Measurement Configuration}

Consider the nonlocal dispersion cancellation experiment shown in Fig.~1.  Here, signal and reference beams propagate through dispersive elements whose dispersion coefficients are equal in magnitude but opposite in sign.  The fields emerging from the dispersive elements illuminate a pair of photodetectors whose photocurrents will be cross correlated to test for dispersion cancellation.  In particular, the signature of nonlocal dispersion cancellation is that this photocurrent cross correlation has a peak whose width is the same with or without the presence of the dispersive elements in the signal and reference paths.
\begin{figure}[h]
\begin{center}
\includegraphics[width=3.25in]{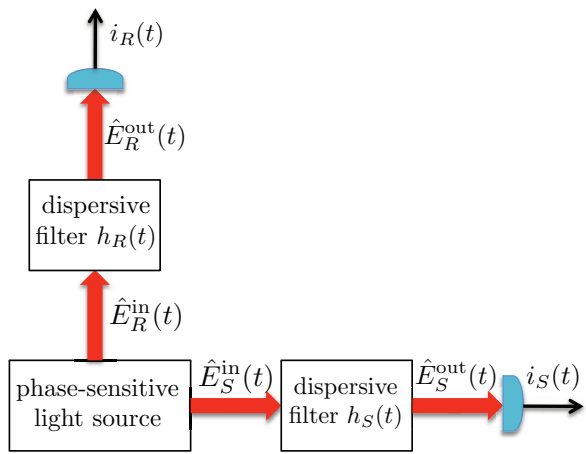}
\end{center}
\caption{(Color online) Configuration for nonlocal dispersion cancellation.  The light source produces signal and reference fields with a common center frequency $\omega_0$ and baseband field operators $\hat{E}_S(t)$ and $\hat{E}_R(t)$, respectively.  The joint signal-reference state has a non-zero phase-sensitive cross correlation.  These fields pass through linear, time-invariant, dispersive filters---with baseband impulse responses $h_S(t)$ and $h_R(t)$---after which they are photodetected.  The resulting photocurrents, $i_S(t)$ and $i_R(t)$, are subsequently cross correlated (in apparatus that is not shown) to seek a signature for nonlocal dispersion cancellation.}
\end{figure}

For simplicity, we will suppress the spatial and polarization characteristics of the signal and reference beams, treating them as time-dependent, scalar, positive-frequency, $\sqrt{\mbox{photons/s}}$-units field operators, $\hat{E}_S(t)e^{-i\omega_0t}$ and $\hat{E}_R(t)e^{-i\omega_0t}$, respectively \cite{footnote0}, with a common center frequency $\omega_0$ and the usual $\delta$-function commutator brackets for their baseband field operators,
\begin{eqnarray}
[\hat{E}_J^{\rm in}(t), \hat{E}_K^{\rm in}(u)] &=& 0 \\[.05in] 
[\hat{E}_J^{\rm in}(t), \hat{E}_K^{{\rm in}\dagger}(u)] &=& \delta_{JK}\delta(t-u),
\end{eqnarray}
for $J,K = S,R$.
The baseband fields operators that these inputs produce at the output of the dispersive elements are then
\begin{equation}
\hat{E}_K^{\rm out}(t) = \int\!du\,\hat{E}_K^{\rm in}(u)h_K(t-u),\mbox{ for $K = S,R$},
\end{equation}
where 
\begin{equation}
h_K(t) = \int\!\frac{d\omega}{2\pi}\,H_K(\omega)e^{i\omega t},
\end{equation}
gives the baseband impulse response of the dispersive element in the signal ($K = S$) or reference ($K=R$) path in terms of its associated frequency response 
\begin{equation}
H_K(\omega) = e^{i\omega_0\tau_p} e^{-i(\omega\tau_g  + \omega^2\beta_K)},
\end{equation}
with $\tau_p$ and $\tau_g$ being its phase and group delays and $\beta_K$ its dispersion coefficient \cite{footnote1}.  In keeping with the usual construct for nonlocal dispersion cancellation, we assume that $\beta_S = - \beta_R = \beta \neq 0$.  Because the dispersive filters are lossless, commutator-bracket preservation is ensured without the need for additional quantum noise, viz., we have that
\begin{eqnarray}
[\hat{E}_J^{\rm out}(t), \hat{E}_K^{\rm out}(u)] &=& 0 \\[.05in] 
[\hat{E}_J^{\rm out}(t), \hat{E}_K^{{\rm out}\dagger}(u)] &=& \delta_{JK}\delta(t-u), 
\label{commutator}
\end{eqnarray}
for $J,K = S,R$.

The photodetectors in Fig.~1 produce classical photocurrents, $i_K(t)$ for $K=S,R$, whose measurement statistics are equivalent to those of the photocurrent operators,
\begin{equation}
\hat{i}_K(t) \equiv q\int\!du\hat{E}_K'^\dagger(u)\hat{E}_K'(u)g(t-u),\mbox{ for $K = S,R$,}
\label{photocurrent}
\end{equation}
where $q$ is the electron charge, 
\begin{equation}
\hat{E}_K'(t) \equiv \sqrt{\eta}\,\hat{E}_K^{\rm out}(t) + \sqrt{1-\eta}\,\hat{E}_{\eta_K}(t),
\label{vacuumnoise}
\end{equation}
with the $\{\hat{E}_{\eta_K}(t)\}$ being baseband field operators that are in their vacuum states, and $0 < \eta \le 1$ is the detector quantum efficiency \cite{footnote2}.  The real-valued function $g(t)$ is the photodetectors' baseband impulse response, which obeys the normalization condition
\begin{equation}
\int\!dt\,g(t) = 1.
\end{equation}

The photocurrents from the two photodetectors are processed in a time-average cross correlator to yield an estimate of the ensemble-average cross correlation
\begin{equation}
C(\tau) \equiv \langle \hat{i}_S(t+\tau)\hat{i}_R(t)\rangle,
\label{crosscorrdefn}
\end{equation}
where our notation anticipates the fact that the joint signal-reference states we shall consider will lead to a cross-correlation function that only depends on the time difference between the photocurrent time samples.  Also, for the purposes of this paper, it suffices to focus on the ensemble average behavior, because the dispersion cancellation effect we are seeking presents its signature there.  Note that the classical photocurrents $\{i_K(t)\}$ associated with measurement of the $\{\hat{i}_K(t)\}$ take the form
\begin{equation}
i_K(t) = q\sum_n g(t-t_{K_n}),
\label{photocounts}
\end{equation}
where the $\{t_{K_n}\}$ are the times at which the signal ($K = S$) or the reference ($K=R$) detector emits a charge carrier in response to its illumination.  When the joint signal-reference field state has sufficiently low photon flux in each beam, the preceding photocurrents will consist of non-overlapping pulses representing individual photon detections, i.e., operation is in the photon counting regime.

\section{Photocurrent Cross Correlation for Gaussian Inputs}

In all that follows we shall restrict our attention to cases in which the joint signal-reference state produced by the source block in Fig.~1 is a zero-mean, continuous-wave, jointly Gaussian state that is completely characterized by the following non-zero correlation functions:  their normally-ordered (phase-insensitive) autocorrelation functions,
\begin{equation}
K_{KK}^{{\rm in}(n)}(\tau) \equiv \langle \hat{E}_K^{{\rm in}\dagger}(t+\tau)\hat{E}_K^{\rm in}(t)\rangle,
\mbox{ for $K = S,R$,}
\label{autocovar}
\end{equation}
and their phase-sensitive cross-correlation function,
\begin{equation} 
K_{SR}^{{\rm in}(p)}(\tau) \equiv \langle \hat{E}_S^{\rm in}(t+\tau)\hat{E}_R^{\rm in}(t)\rangle.
\label{crosscovar}
\end{equation}
These stationary correlation functions have associated spectra \cite{footnote3} given by 
\begin{equation}
{\cal{S}}^{{\rm in}(n)}_{KK}(\omega) \equiv \int\!d\tau\, K_{KK}^{{\rm in}(n)}(\tau)e^{i\omega \tau},
\end{equation}
and
\begin{equation}
{\cal{S}}^{{\rm in}(p)}_{SR}(\omega) \equiv \int\!d\tau\, K_{SR}^{{\rm in}(p)}(\tau)e^{i\omega \tau},
\end{equation}
which will be of use in determining the correlations of the output field operators.  As shown in \cite{Kim}, proper choice of the preceding correlation functions yields the correct quantum statistics for single-spatial-mode outputs from a continuous-wave spontaneous parametric downconverter in the absence of pump depletion.

Because zero-mean Gaussian states with stationary correlations are closed under linear time-invariant transformations, we have that the joint signal-reference state at the output of the dispersive elements in Fig.~1 is also a zero-mean Gaussian state that is completely characterized by its non-zero correlation functions, which are
\begin{eqnarray}
K_{KK}^{{\rm out}(n)}(\tau) &\equiv& \langle \hat{E}_K^{{\rm out}\dagger}(t+\tau)\hat{E}_K^{\rm out}(t)\rangle \\[.05in]
&=& \int\!\frac{d\omega}{2\pi}\,{\cal{S}}_{KK}^{{\rm out}(n)}(\omega)e^{-i\omega\tau} \\[.05in]
&=& \int\!\frac{d\omega}{2\pi}\,{\cal{S}}_{KK}^{{\rm in}(n)}(\omega)|H_{K}(\omega)|^2e^{-i\omega\tau}\\[.05in]
&=& \int\!\frac{d\omega}{2\pi}\,{\cal{S}}_{KK}^{{\rm in}(n)}(\omega)e^{-i\omega\tau}\\[.05in]
&=& K_{KK}^{{\rm in}(n)}(\tau),
\label{KkkOUT}
\end{eqnarray}
and
\begin{eqnarray}
\lefteqn{K_{SR}^{{\rm out}(p)}(\tau) \equiv \langle \hat{E}_S^{{\rm out}}(t+\tau)\hat{E}_R^{\rm out}(t)\rangle} \\[.05in]
&=& \int\!\frac{d\omega}{2\pi}\,{\cal{S}}_{SR}^{{\rm out}(p)}(\omega)e^{-i\omega\tau} \\[.05in]
&=& \int\!\frac{d\omega}{2\pi}\,{\cal{S}}_{SR}^{{\rm in}(p)}(\omega)H_{S}(-\omega)H_{R}(\omega)e^{-i\omega\tau}\\[.05in]
&=& \int\!\frac{d\omega}{2\pi}\,{\cal{S}}_{SR}^{{\rm in}(p)}(\omega)e^{-i\omega^2(\beta_S+\beta_R)}e^{-i\omega\tau}
\label{dispersion}\\[.05in]
&=& \int\!\frac{d\omega}{2\pi}\,{\cal{S}}_{SR}^{{\rm in}(p)}(\omega)e^{-i\omega\tau} \\[.05in]
&=& K_{SR}^{{\rm in}(p)}(\tau).
\label{KsrOUT}
\end{eqnarray}

Equations~(\ref{KkkOUT}) and (\ref{KsrOUT}) embody nonlocal dispersion cancellation for \em both\/\rm\ quantum and classical Gaussian states with phase-sensitive cross correlations.  This is because zero-mean, continuous-wave Gaussian states are completely characterized by their nonzero correlation functions.  Suppose, as we have assumed in this section, that the nonzero correlation functions at the input to the dispersive elements are $K_{KK}^{{\rm in}(n)}(\tau)$, for $K = S,R$, and $K_{SR}^{{\rm in}(p)}(\tau)$.  Further suppose, as we have assumed Sec.~II, that the dispersive elements have identical phase delays, identical group delays, and dispersion coefficients that are equal in magnitude and opposite in sign.  Then, as we have just shown, the nonzero correlation functions at the outputs of the dispersive elements---$K_{KK}^{{\rm out}(n)}(\tau)$, for $K = S,R$, and $K_{SR}^{{\rm out}(p)}(\tau)$---coincide with their counterparts at the inputs to the dispersive elements.  Consequently the state of---i.e., the joint density operator for---the output fields is the same as the state of the input fields.  Inasmuch as the signal and reference fields encounter dispersive elements that do \em not\/\rm\ change their joint state, it is certainly appropriate to say the dispersion has been cancelled in the Fig.~1 setup.  Moreover, because the signal and reference fields encounter spatially-separated dispersive elements and the resulting output fields do not interact or interfere with each other prior to their being photodetected, it is certainly appropriate to say that this dispersion cancellation is a nonlocal effect.   Finally, we note that this state preservation---and hence the nonlocal dispersion cancellation---occurs regardless of whether the input state is classical or quantum, i.e., regardless of whether it has a proper $P$ representation or does not,  To make this completely explicit, for the dispersion cancellation measurement in the Fig.~1 setup, let us use our correlation-function results to evaluate $C(\tau)$.  

Starting from Eqs.~(\ref{photocurrent}) and (\ref{crosscorrdefn}), we find that
\begin{eqnarray}
\lefteqn{C(\tau) =  }\nonumber \\[.05in]
&& q^2\int\!du\!\int\!dv\, \langle\hat{E}_S'^\dagger(u)\hat{E}_S'(u)\hat{E}_R'^\dagger(v)\hat{E}_R'(v)\rangle \nonumber \\[.05in]
&& \times\,\,g(t+\tau-u)g(t-v)\\[.05in]
&=& q^2\eta^2\int\!du\!\int\!dv\,\langle\hat{E}_S^{{\rm out} \dagger}(u)\hat{E}_R^{{\rm out}\dagger}(v)
\hat{E}_S^{\rm out}(u)\hat{E}_R^{\rm out}(v)\rangle \nonumber \\[.05in]
&& \times\,\,g(t+\tau-u)g(t-v)
\label{commutatorconseq}\\[.05in]
&=& q^2\eta^2\int\!du\int\!dv\,(\langle\hat{E}_S^{{\rm out} \dagger}(u)\hat{E}_S^{\rm out}(u)\rangle\langle\hat{E}_R^{{\rm out}\dagger}(v)\hat{E}_R^{\rm out}(v)\rangle \nonumber \\[.05in]
&&+\,\, |\langle \hat{E}_S^{\rm out}(u)\hat{E}_R^{\rm out}(v)\rangle|^2)g(t+\tau-u)g(t-v),
\label{momentfactor}
\end{eqnarray}
where Eq.~(\ref{commutatorconseq}) follows from Eqs.~(\ref{commutator}), and (\ref{vacuumnoise}), and Eq.~(\ref{momentfactor}) follows from the Gaussian-state moment factoring theorem plus our assumption that the joint signal-reference state is zero mean with no phase-insensitive cross correlation \cite{ShapiroSun}.  Using our results for the output field-operators' correlations, Eq.~(\ref{momentfactor}) reduces to 
\begin{eqnarray}
\lefteqn{C(\tau) = q^2\eta^2} \nonumber \\[.05in]
&\times& \left[K_{SS}^{{\rm in}(n)}(0)K_{RR}^{{\rm in}(n)}(0)
+ \int\!du\!\int\!dv\,|K_{SR}^{{\rm in}(p)}(u-v)|^2\right. \nonumber \\[.05in]
&& \left.\times\,\,g(t+\tau -u)g(t-v)\right]\\[.05in]
&=& q^2\eta^2\left[K_{SS}^{{\rm in}(n)}(0)K_{RR}^{{\rm in}(n)}(0) \right. \nonumber \\[.05in]
&& +\,\,\left.
\int\!dz\,|K_{SR}^{{\rm in}(p)}(z)|^2R_{gg}(\tau-z)\right], 
\label{basicanswer}
\end{eqnarray}
where
\begin{equation}
R_{gg}(\tau) \equiv \int\!dt\,g(t+\tau)g(t)
\end{equation}
is the autocorrelation integral of the photodetectors' impulse response $g(t)$.

The first term in Eq.~(\ref{basicanswer}) is rightfully termed the accidental coincidences, inasmuch as it would still be present were there no correlation between the Gaussian states of the signal and reference beams.  It is the second term in which nonlocal dispersion cancellation occurs.  This is  because: (1) it comes from the phase-sensitive cross correlation between the \em output\/\rm\ signal and reference beams; (2) each individual output has encountered a different dispersive element, because $\beta_K \neq 0$ for $K = S,R$ with $\beta_S \neq \beta_R$; and (3) this term does \em not\/\rm\ suffer any dispersion, because $K_{SR}^{{\rm out}(p)}(\tau) = K_{SR}^{{\rm in}(n)}(\tau)$ when $\beta_S = -\beta_R = \beta \neq 0$.  Note that the derivation of Eq.~(\ref{basicanswer}) \em only\/\rm\ assumes that the joint signal-reference state at the input to the dispersive elements in Fig.~1 is zero-mean and Gaussian with non-zero correlations given by Eqs.~(\ref{autocovar}) and (\ref{crosscovar}).  Thus, it applies to \em both\/\rm\ quantum \em and\/\rm\ classical states, by appropriate choice of these correlations.  Furthermore, although we have used quantum notation in our derivation of Eq.~(\ref{basicanswer}), the same result would be obtained for classical-state light if we used the semiclassical theory of photodetection as follows \cite{footnote0}.  (1)  We assume that the baseband signal and reference fields at the input to the dispersive elements in Fig.~1 are zero-mean, jointly-Gaussian classical random processes, $E_S^{\rm in}(t)$ and $E_R^{\rm in}(t)$,  that are completely characterized by their non-zero correlations 
\begin{equation}
K_{KK}^{{\rm in}(n)}(\tau) \equiv \langle E_K^{{\rm in}*}(t+\tau)E_K^{\rm in}(t)\rangle,
\mbox{ for $K = S,R$,}
\label{Cautocovar}
\end{equation}
and \begin{equation} 
K_{SR}^{{\rm in}(p)}(\tau) \equiv \langle E_S^{\rm in}(t+\tau)E_R^{\rm in}(t)\rangle.
\label{Ccrosscovar}
\end{equation} 
(2)  We calculate the photocurrent statistics by assuming the event times $\{t_{K_n}\}$ comprise independent Poisson point processes, for $K=S,R$, conditioned on knowledge of the fields illuminating the photodetectors, and that the conditional rate functions for these Poisson point processes are 
\begin{equation}
\mu_K(t) = \eta|E_K^{\rm out}(t)|^2, \mbox{ for $K = S,R$.}
\end{equation}
In the next section we will instantiate Eq.~(\ref{basicanswer}) in two special cases of zero-mean, continuous-wave, jointly Gaussian states.  In the first, the input signal and reference fields have the maximum phase-sensitive cross correlation permitted by quantum mechanics, i.e., they are in a nonclassical state.  In the second, the joint signal-reference state has a phase-sensitive cross correlation that is at the tighter limit set by classical physics.  Hence, it is has a proper $P$ representation and is thus a classical state.  

\section{Quantum versus Classical-State Dispersion Cancellation}
Suppose that the signal and reference correlation functions at the input to the dispersive elements in Fig.~1 are as follows:
\begin{equation}
K_{SS}^{{\rm in}(n)}(\tau) = K_{RR}^{{\rm in}(n)}(\tau) = K^{{\rm in}(n)}(\tau) \equiv Pe^{-\tau^2/2T_0^2},
\label{normalorder}
\end{equation}
and $K_{SR}^{{\rm in}(p)}(\tau) = K_{SR}^{(q)}(\tau)$ \em or\/\rm\ $K_{SR}^{{\rm in}(p)}(\tau) = K_{SR}^{{\rm in}(c)}(\tau)$, where 
\begin{equation}
K_{SR}^{{\rm in}(q)}(\tau) \equiv Pe^{-\tau^2/2T_0^2} + i \sqrt{P}\!\left(\frac{2}{\pi T_0^2}\right)^{1/4}\!e^{-\tau^2/T_0^2},
\label{quantumCorr}
\end{equation}
and 
\begin{equation}
K_{SR}^{{\rm in}(c)}(\tau) \equiv Pe^{-\tau^2/2T_0^2}.
\label{classicalCorr}
\end{equation}
The superscripts $(q)$ and $(c)$ denote quantum and classical states respectively, as the following discussion will justify.  Before doing so, however, there is an important point to be made.  Because we are assuming that the signal and reference fields are in a zero-mean, jointly-Gaussian state, then---regardless of their phase-sensitive cross-correlation function---their reduced density operators are zero-mean Gaussian states.  So, because the quantum and classical signal-reference states in this section have the same autocorrelations, there is no single-beam (signal only or reference only) measurement that can distinguish between them.  It is \em only\/\rm\ when joint measurements are made on the signal and reference beams---e.g., the photocurrent cross-correlation measurement employed in the dispersion-cancellation experiment from Fig.~1---that any difference can be discerned between these quantum and classical signal-reference states.  With this point in mind, let us review the quantum and classical limits on the cross spectra associated with the preceding cross-correlation functions

The spectra associated with the correlation functions from Eqs.~(\ref{normalorder})--(\ref{classicalCorr}) are
\begin{eqnarray}
S^{{\rm in }(n)}_{SS}(\omega) &=& S^{{\rm in}(n)}_{RR}(\omega) =  S^{{\rm in}(n)}(\omega) 
\nonumber \\[.05in]
&=& P\sqrt{2\pi T_0^2}e^{-\omega^2T_0^2/2}
\label{normspectrum}\\[.05in]
S_{SR}^{{\rm in}(q)}(\omega) &=& P\sqrt{2\pi T_0^2}e^{-\omega^2T_0^2/2} \nonumber \\[.05in] 
&& +\,\, i\sqrt{P}(2\pi T_0^2)^{1/4}e^{-\omega^2T_0^2/4}
\label{qcrossspectrum}\\[.05in]
S_{SR}^{{\rm in}(c)}(\omega) &=& P\sqrt{2\pi T_0^2}e^{-\omega^2T_0^2/2}.
\label{ccrossspectrum}
\end{eqnarray}
Quantum mechanics sets the following bound on $|S_{SR}^{{\rm in}(p)}(\omega)|$ \cite{ShapiroSun,footnote4}, 
\begin{equation}
|S_{SR}^{{\rm in}(p)}(\omega)| \le \sqrt{S^{{\rm in}(n)}_{SS}(\omega)[1+S^{{\rm in}(n)}_{RR}(-\omega)]},
\label{qbound}
\end{equation}
which Eqs.~(\ref{normspectrum}) and (\ref{qcrossspectrum}) saturate, implying that the joint signal-reference Gaussian state with these spectra is maximally entangled in frequency \cite{footnote5}.  On the other hand, Eqs.~(\ref{normspectrum}) and (\ref{ccrossspectrum}) satisfy, with equality, the tighter bound required by classical physics \cite{ShapiroSun}, 
\begin{equation}
|S_{SR}^{{\rm in}(p)}(\omega)| \le \sqrt{S^{{\rm in}(n)}_{SS}(\omega)S^{{\rm in}(n)}_{RR}(-\omega)},
\label{cbound}
\end{equation}
indicating that the the joint signal-reference Gaussian state with these spectra is classical, with the maximum possible phase-sensitive cross correlation.  Indeed, if $E(t)$ is a complex-valued, zero-mean, Gaussian random process with 
\begin{equation}
\langle E(t+\tau)E(t)\rangle = 0
\end{equation}
and 
\begin{equation}
\langle E^*(t+\tau)E(t)\rangle = Pe^{-\tau^2/2T_0^2}
\end{equation}
then the joint signal-reference Gaussian state with $K_{SS}^{{\rm in}(n)}(\tau) = K_{RR}^{{\rm in}(n)}(\tau) = K^{{\rm in}(n)}(\tau)$ and  $K_{SR}^{{\rm in}(p)}(\tau) = K_{SR}^{{\rm in}(c)}(\tau)$ is a classical mixture of continuous-time coherent states $|E_S^{\rm in}(t)\rangle|E_R^{\rm in}(t)\rangle$ in which $E_S(t) = E(t)$ and $E_R(t) = E^*(t)$.  

Using the results of the preceding paragraph in Eq.~(\ref{basicanswer}), in conjunction with the convenient choice 
\begin{equation}
g(t) = \frac{e^{-t^2/T_g^2}}{\sqrt{\pi T_g^2}},
\end{equation}
we find that
\begin{equation}
C^{(c)}(\tau) = q^2\eta^2P^2\!\left(1+ \frac{e^{-\tau^2/(T_0^2 + 2T_g^2)}}{\sqrt{1+ 2T_g^2/T_0^2}}\right),
\label{cSlowDet}
\end{equation}
and 
\begin{equation}
C^{(q)}(\tau) = C^{(c)}(\tau) + q^2\eta^2P\frac{e^{-2\tau^2/(T_0^2 + 4T_g^2)}}{\sqrt{\pi(  T_0^2/2+2T_g^2)}},
\label{qSlowDet}
\end{equation}
with the superscripts distinguishing between the quantum and classical-state cases. 
In both of these expressions the constant term $q^2\eta^2P^2$ comes from the accidental coincidences noted earlier.  Thus we see that the contrast between the dispersion-cancellation terms and the accidental coincidences degrades for $T_g \gg T_0$, i.e., when the photodetectors' response time is long compared to the coherence time of the signal and reference.  So, to best understand the difference between the quantum and classical cases, let us assume we have detectors that are fast enough to yield 
\begin{equation}
C^{(c)}(\tau) \approx q^2\eta^2P^2(1+ e^{-\tau^2/T_0^2}),
\label{classicalanswer}
\end{equation}
and
\begin{eqnarray}
C^{(q)}(\tau) &\approx& q^2\eta^2P^2(1+ e^{-\tau^2/T_0^2}) \nonumber \\[.05in]
&& +\,\, q^2\eta^2Pe^{-2\tau^2/T_0^2}/\sqrt{\pi T_0^2/2} \\[.05in]
&\approx& q^2\eta^2P^2\!\left(1+ \frac{e^{-2\tau^2/T_0^2}}{PT_0\sqrt{\pi/2}}\right),
\label{quantanswer}
\end{eqnarray}
where we have used the low-brightness condition $PT_0 \ll 1$ \cite{footnote6} to obtain (\ref{quantanswer}).

Comparison of Eqs.~(\ref{classicalanswer}) and\,(\ref{quantanswer}) reveal that both of these photocurrent cross correlations consist of the \em same\/\rm\ background term, $C_{\rm acc} \equiv q^2\eta^2P^2$, arising from accidental coincidences, plus a Gaussian-shaped term that is the signature of the non-zero phase-sensitive cross correlation between the signal and reference fields.  In both cases this signature term enjoys dispersion cancellation, because it is independent of the non-zero value of the dispersion coefficients, $\beta_S = -\beta_R = \beta \neq 0$.  Moreover, Eqs.~(\ref{dispersion}) and (\ref{momentfactor}) imply that \em both\/\rm\ the classical \em and\/\rm\ the quantum signature terms would increasingly broaden from dispersion, for $\beta_S \neq -\beta_R$, as $|\beta_S + \beta_R|$ grows without bound.  What then are the differences between $C^{(c)}(\tau)$ and $C^{(q)}(\tau)$ in this fast-detector, low-brightness regime?  There are two.  First, as we have previously found for a comparable spatial case in ghost imaging \cite{Erkmen}, the width of the dispersion-cancelled signature term for the quantum case $C_{\rm dc}^{(q)}(\tau)$ is different from that of the corresponding classical case $C_{\rm dc}^{(c)}(\tau)$, despite the individual signal and reference fields having the same fluorescence bandwidths in both instances \cite{footnote8}.  Second, and more significantly, the contrast between the dispersion-cancelled quantum term $C_{\rm dc}^{(q)}(\tau)$ and the accidentals term $C_{\rm acc}$, given by
\begin{equation}
{\cal{C}}^{(q)} \equiv \max_\tau\frac{C_{\rm dc}^{(q)}(\tau)}{C_{\rm acc}}  \approx \frac{1}{PT_0\sqrt{\pi/2}} \gg 1,
\end{equation}
dramatically exceeds that for the classical-state case
\begin{equation}
{\cal{C}}^{(c)} \equiv \max_\tau\frac{C_{\rm dc}^{(c)}(\tau)} 
{C_{\rm acc}}  \approx 1.
\end{equation}
This too is a feature that has been seen in comparing quantum and classical-state versions of ghost imaging \cite{Erkmen}.

\section{Discussion}
We have applied Gaussian-state analysis to a version of Franson's nonlocal dispersion cancellation paradigm.  In the fast-detector regime using a low-brightness source of signal and reference beams with phase-sensitive cross correlation we showed that both quantum (maximally entangled) and classical-state (maximally correlated) sources produced ensemble-average photocurrent cross correlations comprised of a constant background term, arising from accidental coincidences, plus a dispersion-cancelled signature term.  The signature-term widths obtained with the classical and nonclassical sources are different, for Gaussian fluorescence spectra of the same bandwidth, but this is not an essential feature \cite{footnote8}.  The major difference between these two cases is in their contrast.  The quantum source yields very high contrast ($\gg$1) dispersion cancellation, while the classical-state source has a contrast equal to 1.  Nevertheless, both dispersion-cancelled signatures---quantum and classical---arise from the propagation of a phase-sensitive cross correlation through the dispersive elements in the signal and reference paths, i.e., their physical origins are identical and essentially classical.  It is the greatly enhanced observability of the quantum case---which persists well into the slow-detector ($T_g \gg T_0$) regime at low source brightness---that really distinguishes it from its classical counterpart.  Indeed, for reasonable experimental parameters for a downconverter source and single-photon detection system---$P = 10^6$\,pairs/s, $T_0 = 1$\,ps, and $T_g =1$\,ns---we find that 
\begin{equation}
{\cal{C}}^{(q)} \approx \frac{1}{\sqrt{2\pi}\, PT_g} \approx 399,
\end{equation}
whereas 
\begin{equation}
{\cal{C}}^{(c)} \approx \frac{T_0}{\sqrt{2}\,T_g} \approx 7 \times 10^{-4} .
\end{equation}
Inasmuch as this low-brightness, slow-detector regime is the norm for downconverter coincidence counting---including dispersion-cancellation experiments---these contrast values show the dramatic benefit of having a quantum, rather than a classical-state, source available. 

As final elaboration on the conclusions reached in the preceding paragraph, we shall discuss two additional limits of our Gaussian-state analysis for the quantum signal-reference state, plus a culminating example illustrating a smooth transition from quantum to classical-state sources.  The first limiting case is low-flux operation, which will connect our work for the quantum case to the more frequently employed biphoton treatment.  The second limiting case is high-brightness operation, which will link our work for the quantum case to the results we obtained for the classical signal-reference state.  The final example uses the bandlimited spectra specified in \cite{footnote8}, in conjunction with additive noise, to study contrast degradation in the dispersion-cancelled photocurrent cross correlation as the input signal-reference state is continuously varied from maximally entangled to maximally correlated to partially correlated to uncorrelated.  

\subsection{Low-Flux Operation}
Consider the single-spatial-mode signal ($S$) and idler ($I$) outputs from a frequency-degenerate continuous-wave parametric downconverter.  In the absence of pump depletion, they are in a zero-mean jointly Gaussian state that is completely characterized by the non-zero correlation functions of the associated baseband field operators, namely
\begin{equation}
K_{KK}^{(n)}(\tau) \equiv \langle \hat{E}_K^\dagger(t+\tau)\hat{E}_K(t)\rangle,
\mbox{ for $K = S,I$,}
\end{equation}
and 
\begin{equation} 
K_{SI}^{(p)}(\tau) \equiv \langle \hat{E}_S(t+\tau)\hat{E}_I(t)\rangle.
\end{equation} 
For type-II phase matching with a timing-compensation crystal employed at the downconverter's output, the spectra associated with these correlation functions in the low-brightness regime are \cite{Kim}
\begin{equation}
S_{KK}^{(n)}(\omega) = (\gamma |E_P|\ell)^2 \!\left(\frac{\sin(\omega \Delta k' \ell/2)}{\omega \Delta k'\ell/2}\right)^2,
\end{equation}
and
\begin{equation}
S_{SI}^{(p)}(\omega) = i \gamma E_P\ell\frac{\sin(\omega \Delta k'\ell/2)}{\omega \Delta k'\ell/2},
\end{equation}
where $\gamma$ is the nonlinear coefficient from the coupled-mode equations, $E_P$ is the classical baseband phasor for the pump field, $\ell$ is the crystal length, and $\Delta k'$ is the phase mismatch coefficient, i.e., $\omega\Delta k'$ is the phase mismatch at detuning $\omega$ from frequency degeneracy.  When the source flux is low enough that $K_{SS}^{(n)}(0)T = K_{II}^{(n)}(0)T \ll 1$, where $T \gg T_g$ is the maximum $|\tau|$ for which we are trying to estimate the ensemble-average photocurrent cross correlation $C(\tau)$, we can neglect multiple-pair emissions.  Hence the jointly Gaussian state of the signal and idler can be taken to be a predominant vacuum term plus a weak biphoton component \cite{Kim}, viz., 
\begin{eqnarray}
\lefteqn{|\psi\rangle_{SI} \approx |{\bf 0}\rangle_S|{\bf 0}\rangle_I + i\gamma E_P\ell} \nonumber \\[.05in] &\times& \!\!
\int\!\frac{d\omega}{2\pi}\,\frac{\sin(\omega \Delta k'\ell/2)}{\omega \Delta k'\ell/2}|\omega_P/2 + \omega\rangle_S|\omega_P/2 -\omega\rangle_I.
\label{biphotonapprox}
\end{eqnarray}
Here, $\omega_P$ is the pump frequency, $|{\bf 0}\rangle_K$ denotes the multi-mode vacuum state, and $|\omega_P/2 \pm \omega\rangle_K$ denotes a single photon state of the signal ($K=S$) or idler ($K=I$) at frequency $\omega_P/2 \pm \omega$.  Replacing the sinc phase-matching function with the Gaussian approximation to its main lobe \cite{Eberly} will then lead to an ensemble-average photocurrent cross correlation equal to the dispersion-cancelled signature term from Eq.~(\ref{quantanswer}) without any background, once the proper identifications have been made for $P$ and $T_0$ \cite{footnote9}.  Note that the absence of the background term in the biphoton analysis is due to that treatment's neglecting the multiple-pair contributions that are present in the full Gaussian-state characterization of the downconverter's output.  Also note that the close connection between biphoton analysis and our Gaussian-state approach---in this and more general quantum imaging scenarios---follows from the fact that the biphoton wave function propagates according to the same transformation rule as the phase-sensitive cross-correlation function cf.\ \cite{Erkmen} and \cite{BU}.

\subsection{High-Brightness Operation}
Here we turn to what happens to dispersion cancellation when the source in Fig.~1 operates at high brightness.  Specifically, let us revisit the behavior of the photocurrent cross-correlation functions found in Sec.~IV when the quantum and classical signal-reference Gaussian states have the spectra given in Eqs.(\ref{normspectrum})--(\ref{ccrossspectrum}) but satisfying the high-brightness condition, $PT_0 \gg 1$, instead of the low-brightness condition, $PT_0 \ll 1$.  At high source brightness the photocurrent cross correlation for the classical signal-reference state is still given by Eq.~(\ref{cSlowDet}), for arbitrary $T_g$ and $T_0$.   For the quantum signal-reference state, on the other hand, Eq.~(\ref{qSlowDet}) still applies, but the high-brightness condition reduces it to 
\begin{equation}
C^{(q)}(\tau) 
\approx C^{(c)}(\tau),
\label{qSlowDet2}
\end{equation}
indicating that both the quantum and classical signal-reference states give virtually \em identical\/\rm\ dispersion-cancelled photocurrent cross correlations.  This occurs because at high source brightness the difference between the quantum and classical bounds on the phase-sensitive cross spectrum disappears, cf.\@ Eqs.~(\ref{qbound}) and (\ref{cbound}).  Note, however, that the high-brightness quantum state is extremely nonclassical:  combining its signal and reference beams on a 50-50 beam splitter will result in outputs that exhibit very strong quadrature-noise squeezing \cite{ShapiroSun}.  The photocurrent cross-correlation measurement is \em not\/\rm\ sensitive to that effect, hence the high-brightness quantum state looks classical in the Fig.~1 experiment.  Furthermore, high-brightness operation when $T_g \gg T_0$ violates the low-flux condition under which the photocurrents from Eq.~(\ref{photocounts}) contain easily resolvable individual charge-carrier emissions.  Thus in high-brightness operation the photocurrent cross-correlation measurement will no longer correspond to photon-coincidence counting \cite{ShapiroSun}.  

\subsection{Dispersion Cancellation with Additive Noise}
Suppose that the signal and reference beams in the Fig.~1 setup are obtained as follows.  A continuous-wave downconverter is used to produce a zero-mean, jointly Gaussian signal-reference state fully characterized by the following nonzero spectra for the baseband field operators of the signal and idler,
\begin{equation}
S_{KK}^{(n)}(\omega) \left\{\begin{array}{ll}
\pi P/\Omega, & \mbox{for $|\omega| \le \Omega$}\\[.05in]
0, & \mbox{otherwise,}
\end{array}
\right.
\end{equation}
for $K = S,I$, and
\begin{equation}
S_{SI}^{(p)}(\omega) \left\{\begin{array}{ll}
\pi P/\Omega + i\sqrt{\pi P/\Omega}, & \mbox{for $|\omega| \le \Omega$}\\[.05in]
0, & \mbox{otherwise.}
\end{array}
\right.
\end{equation}
(These spectra could be obtained, in principle, by passing the output fields from a very broadband downconverter through an ideal passband filter.)  The input fields in Fig.~1 are then obtained by passing the signal and idler through identical transmissivity-$\kappa$ beam splitters followed first by identical phase-insensitive amplifiers with gain $G = 1/\kappa \ge 1$ and minimum (vacuum-state) noise level and then by identical ideal passband filters.  The resulting signal and reference fields will then be in a zero-mean, jointly Gaussian state that is fully characterized by these nonzero spectra for their baseband field operators:  
\begin{equation}
S_{KK}^{{\rm in}(n)}(\omega) \left\{\begin{array}{ll}
\pi P/\Omega + (G-1), & \mbox{for $|\omega| \le \Omega$}\\[.05in]
0, & \mbox{otherwise,}
\end{array}
\right.
\end{equation}
for $K = S,R$, and
\begin{equation}
S_{SR}^{{\rm in}(p)}(\omega) \left\{\begin{array}{ll}
\pi P/\Omega + i\sqrt{\pi P/\Omega}, & \mbox{for $|\omega| \le \Omega$}\\[.05in]
0, & \mbox{otherwise.}
\end{array}
\right.
\end{equation} 
The state-propagation calculation performed in Sec.~III will show, once again, that this joint signal-reference state is preserved when the dispersive elements have identical phase delays, identical group delays, and dispersion coefficients that are equal in magnitude and opposite in sign.  The photocurrent-correlation calculation from Sec.~III now leads to 
\begin{eqnarray}
C(\tau) &=& q^2\eta^2(P + (G-1)\Omega/\pi)^2 \nonumber \\[.05in]
&& + \,\, q^2\eta^2(P^2 + P\Omega/\pi)\!\left(\frac{\sin(\Omega \tau)}{\Omega \tau}\right)^2,
\label{Csmooth}
\end{eqnarray}
in the fast-detector limit.  

Let us explore the behavior of this $C(\tau)$ result as $\kappa$ decreases from one to zero.  For any value of $\kappa$ we have that $C(\tau)$ consists of an accidentals term ${\cal{C}}_{\rm acc} \equiv q^2\eta^2(P+(G-1)\Omega/\pi)^2$, plus a dispersion-cancelled term ${\cal{C}}_{\rm dc}(\tau)$ that---regardless of the downconverter's brightness and the amount of noise injected by the phase-insensitive amplifier---is proportional to $(\sin(\Omega \tau)/\Omega \tau)^2$.  All that remains, therefore is to examine the contrast between the dispersion-cancelled term and the accidentals.  Here we find that
\begin{equation}
{\cal{C}} \equiv \max_\tau\frac{C_{\rm dc}(\tau)}{C_{\rm acc}} = 
\frac{1+\Omega/\pi P}{(1+(G-1)\Omega/\pi P)^2}.
\label{finalcontrast}
\end{equation}
When $\kappa = 1/G = 1$, the joint signal-reference state is a maximally-entangled pure Gaussian state and Eq.~(\ref{finalcontrast}) yields,
\begin{equation}
{\cal{C}} = {\cal{C}}_{\mbox{\scriptsize max-ent}} \equiv 1+ \Omega /\pi P,
\end{equation}
which monotonically decreases from ${\cal{C}}_{\mbox{\scriptsize max-ent}} \gg 1$, at low source brightness to ${\cal{C}}_{\mbox{\scriptsize max-ent}}\approx 1$ at high source brightness.  On the other hand, for any source brightness we see that ${\cal{C}}$ decreases monotonically with decreasing $\kappa$ (increasing $G$).  Moreover, when 
\begin{equation}
G = G_c \equiv 1 + \frac{\pi P}{\Omega}\!\left(\sqrt{1+ \frac{\Omega}{\pi P}}\right),
\end{equation}
we have that $|S_{SR}^{{\rm in }(p)}(\omega)| = \sqrt{S_{SS}^{{\rm in}(n)}(\omega)S_{RR}^{{\rm in}(n)}(-\omega)}$, so that the joint signal-reference state is a maximally-correlated classical Gaussian mixed state.  In this case ${\cal{C}}$ equals the maximally-correlated result, 
\begin{equation}
{\cal{C}} = {\cal{C}}_{\mbox{\scriptsize max-corr}} = 1.
\end{equation}
Further decreases in $\kappa$ (increases in $G$) continue to degrade ${\cal{C}}$ until it goes to zero as $\kappa \rightarrow 0$ ($G\rightarrow \infty$).    

In conclusion, the preceding example undergoes a continuous progression of the joint signal-reference input state---as $\kappa$ decreases and $G$ increases---from a maximally-entangled Gaussian pure state (when $G=1$), to a nonclassical mixed Gaussian state (when $1< G < G_c$), to a maximally-correlated classical Gaussian mixed state (when $G= G_c$), to a classical mixed Gaussian product state (when $G \rightarrow \infty$).  Accompanying this continuous progression of states is the continuous progression of ${\cal{C}}$ from ${\cal{C}}_{\mbox{\scriptsize max-ent}}$ (for $G=1$), to 
${\cal{C}}_{\mbox{\scriptsize max-corr}} < {\cal{C}} < {\cal{C}}_{\mbox{\scriptsize max-ent}}$ (for $1 < G <G_c$) to ${\cal{C}} = {\cal{C}}_{\mbox{\scriptsize max-corr}}$ (for $G = G_c$) to ${\cal{C}}\rightarrow 0$ (for $G \rightarrow \infty$).  Throughout this progression of states and contrasts, the Fig.~1 setup yields a photocurrent cross-correlation function comprised of an accidentals term plus a fixed shape dispersion-cancelled term.  For $0<\epsilon \ll G_c-1$ the experiment requires quantum photodetection to exactly account for its behavior when $G = G_c -\epsilon$, but semiclassical photodetection suffices when $G = G_c +\epsilon$.  Absent a discontinuity in the physical mechanism for dispersion cancellation, when $G$ crosses from $G < G_c$ to $G>G_c$, then the physical explanations for the dispersion-cancelled terms in these two regimes must be the same.  We assert that there is no such discontinuity.  It is state preservation for zero-mean jointly Gaussian states with a phase-sensitive cross correlation---implied by classical coherence-theory propagation of that cross correlation---that is responsible for the nonlocal dispersion cancellation in the Fig.~1 experiment.  

In short, our Gaussian-state analysis supports Franson's assertion from \cite{Franson1}:  there is no classical explanation that can account for \em all\/\rm\ the features of his nonlocal dispersion-cancellation experiment.  However, our work shows that the only intrinsically quantum-mechanical feature in this experiment is the high contrast that is achieved with a maximally-entangled (biphoton) source.

\section*{ACKNOWLEDGEMENT}

The author acknowledges valuable technical discussion with F~N.~C.~Wong.  This work was supported by U.S. Army Research Office MURI Grant No. W911NF-05-1-0197.


\begin{thebibliography}{2}
\bibitem{footnote}In semiclassical photodetection theory the light illuminating the photodetector is a classical (possibly random) electromagnetic wave and the discreteness of the electron charge leads to photodetection shot noise as the fundamental source of noise.
\bibitem{footnote0} We will use quantum notation throughout, to enable a unified treatment of both classical and nonclassical illumination.  The reader should keep in mind that whenever the field states involved are classical, then quantitatively identical predictions for the measurement statistics result from treating the fields as classical and employing semiclassical photodetection theory rather than quantum photodetection theory.  See \cite{Erkmen} for a detailed discussion of this point in the context of quantum versus classical Gaussian states in ghost imaging.
\bibitem{footnote1}Without loss of generality, with respect to dispersion cancellation, we have assumed that both dispersive elements have the same phase delay and the same group delay.
\bibitem{footnote3}Physically, $S_{KK}^{{\rm in}(n)}(\omega)$ is the fluorescence spectrum of $\hat{E}^{\rm in}_K(t)e^{-i\omega_0t}$ at frequency $\omega_0 + \omega$.  As a result, it must be non-negative at all frequencies, whereas the phase-sensitive cross spectrum, $S^{{\rm in}(p)}_{SR}(\omega)$ is, in general, complex valued.
\bibitem{footnote4}That this inequality---and the one to follow for the classical-state source---involves the fluorescence spectra of the signal and reference at $\omega_0 + \omega$ and $\omega_0 - \omega$, respectively, is a manifestation of the bichromatic nature of stationary phase-sensitive  cross correlations.
\bibitem{footnote5}The biphoton approximation to the joint signal-reference Gaussian state, which we will exhibit in Sec.~V, makes this frequency entanglement explicit.
\bibitem{footnote6}Continuous-wave spontaneous parametric downconversion is ordinarily operated in the low-brightness regime, wherein the average number of signal-idler pairs emitted during a coherence time is much less than one \cite{Kim}.
\bibitem{footnote8}The difference in widths between the quantum and classical dispersion-cancelled terms depends on their phase-sensitive cross spectra.  Suppose we have the strictly bandlimited spectra $S_{SS}^{{\rm in}(n)}(\omega) = S_{RR}^{{\rm in}(n)}(\omega) = S_{SR}^{{\rm in}(c)}(\omega) = \pi P/\Omega$ and $S_{SR}^{{\rm in}(q)}(\Omega) = \pi P/\Omega + i\sqrt{\pi P/\Omega}$ for $|\omega|\le \Omega$ and zero otherwise.  We then find that:  (1)  the quantum cross spectrum is at its ultimate (maximally-entangled) limit $|S_{SR}^{{\rm in}(q)}(\omega)| = \sqrt{S_{SS}^{{\rm in}(n)}(\omega)[1+S_{RR}^{{\rm in}(n)}(-\omega)]}$; (2) the classical cross spectrum is at its ultimate (maximally-correlated) limit $|S_{SR}^{{\rm in}(c)}(\omega)| = \sqrt{S_{SS}^{{\rm in}(n)}(\omega)S_{RR}^{{\rm in}(n)}(-\omega)}$; and (3) the photocurrent cross correlations only differ in their contrast, i.e., they have identical accidentals $q^2\eta^2P^2$, and their dispersion-cancelled terms are both proportional to $(\sin(\Omega \tau)/\Omega \tau)^2$.   See Sec.~V.C for more details.
\end{thebibliography}
\end{document}